\documentstyle[twocolumn,aps,floats,amssymb,epsfig]{revtex}

\begin{document}

\def\d{{\rm d}}
\def\e{{\rm e}}
\def\O{{\rm O}}
\def\half{\mbox{$\frac12$}}
\def\eref#1{(\protect\ref{#1})}
\def\etal{{\it{}et~al.}}

\draft
\tolerance = 10000

\renewcommand{\topfraction}{0.9}
\renewcommand{\textfraction}{0.1}
\renewcommand{\floatpagefraction}{0.9}

\twocolumn[\hsize\textwidth\columnwidth\hsize\csname @twocolumnfalse\endcsname

\title{The structure of scientific collaboration networks}
\author{M. E. J. Newman}
\address{Santa Fe Institute, 1399 Hyde Park Road, Santa Fe, NM 87501}
\maketitle

\begin{abstract}
  We investigate the structure of scientific collaboration networks.  We
  consider two scientists to be connected if they have authored a paper
  together, and construct explicit networks of such connections using data
  drawn from a number of databases, including MEDLINE (biomedical
  research), the Los Alamos e-Print Archive (physics), and NCSTRL (computer
  science).  We show that these collaboration networks form ``small
  worlds'' in which randomly chosen pairs of scientists are typically
  separated by only a short path of intermediate acquaintances.  We further
  give results for mean and distribution of numbers of collaborators of
  authors, demonstrate the presence of clustering in the networks, and
  highlight a number of apparent differences in the patterns of
  collaboration between the fields studied.
\end{abstract}

\pacs{}

]

\vspace{1cm}

\section{Introduction}
A {\bf social network} is a collection of people, each of whom is
acquainted with some subset of the others.  Such a network can be
represented as a set of points (or {\bf vertices}) denoting people, joined
in pairs by lines (or {\bf edges}) denoting acquaintance.  One could, in
principle, construct the social network for a company or firm, for a school
or university, or for any other community up to and including the entire
world.

Social networks have been the subject of both empirical and theoretical
study in the social sciences for at least fifty
years~\cite{Kochen89,WF94,Valente95,Watts99}, partly because of inherent
interest in the patterns of human interaction, but also because their
structure has important implications for the spread of information and
disease.  It is clear for example that variation in just the average number
of acquaintances which individuals have (also called the average {\bf
  degree} of the network) might substantially influence the propagation of
a rumor, a fashion, a joke, or this year's flu.

One of the first empirical studies of the structure of social networks,
conducted by Stanley Milgram~\cite{Milgram67}, asked test subjects, chosen
at random from a Nebraska telephone directory, to get a letter to a target
subject in Boston, a stockbroker friend of Milgram's.  The instructions
were that the letters were to be sent to their addressee (the stockbroker)
by passing them from person to person, but that they could be passed only
to someone whom the passer knew on a first-name basis.  Since it was not
likely that the initial recipients of the letters were on a first-name
basis with a Boston stockbroker, their best strategy was to pass their
letter to someone whom they felt was nearer to the stockbroker in some
sense, either social or geographical: perhaps someone they knew in the
financial industry, or a friend in Massachusetts.

A moderate number of Milgram's letters did eventually reach their
destination, and Milgram discovered that the average number of steps taken
to get there was only about six, a result which has since passed into
folklore and was immortalized by John Guare in the title of his 1990 play
{\it Six Degrees of Separation}~\cite{Guare90}.  Although there were
certainly biases present in Milgram's experiment---letters which took a
longer path were perhaps more likely to get lost or forgotten, for
instance---Milgram's result is usually taken as evidence of the ``small
world hypothesis,'' that most pairs of people in a population can be
connected by only a short chain of intermediate acquaintances, even when
the size of the population is very large.

Milgram's work, although cleverly conducted and in many ways revealing,
does not however tell us much about the detailed structure of social
networks, data that are crucial to the understanding of information or
disease propagation.  Many subsequent studies have addressed this problem.
Discussions can be found in Refs.~\onlinecite{WF94}
and~\onlinecite{Watts99}.  A number of investigations of real acquaintance
networks have been performed.  Foster~\etal~\cite{FRO63} and Fararo and
Sunshine~\cite{FS64}, for instance, both constructed maps of friendship
networks among high-school students, and Bernard~\etal\ did the same for
communities of Utah Mormans, Native Americans, and Micronesian
islanders~\cite{BKEMS98}.  Surveys or interviews were used to determine
friendships.  While these studies directly probe the structure of the
relevant social network, they suffer from two substantial shortcomings
which limit their usefulness.  First, the studies are labor-intensive and
the size of the network which can be mapped is therefore
limited---typically to a few tens or hundreds of people.  Second, these
studies are highly sensitive to subjective bias on the part of
interviewees; what is considered to be an ``acquaintance'' can differ
considerably from one person to another.

To avoid these issues, a number of researchers have studied networks for
which there exist more numerous data and more precise definitions of
connectedness.  Examples of such networks are the electric power
grid~\cite{WS98}, the internet~\cite{AJB99,BKMRRSTW00}, and the pattern of
air traffic between airports~\cite{ASBS00}.  These networks however suffer
from a different problem: although they may loosely be said to be social
networks in the sense that their structure in some way reflects features of
the society which built them, they do not directly measure actual contact
between people.  Many researchers are, of course, interested in these
networks for their own sake, but to the extent that we want to know about
human acquaintance patterns, power grids and computer networks are a poor
proxy for the real thing.

Perhaps the nearest that studies of this kind have come to looking at a
true acquaintance network is in studies of the network of movie
actors~\cite{WS98,ASBS00}.  In this network, which has been thoroughly
documented and contains nearly half a million people, two actors are
considered connected if they have been credited with appearance in the same
film.  However, while this is genuinely a network of people, it is far from
clear that the appearance of two actors in the same movie implies that they
are acquainted in any but the most cursory fashion, or that their
acquaintance extends off-screen.  To draw conclusions about patterns of
every-day human interaction from the movies would, it seems certain, be a
mistake.

In this paper, we present a study of a genuine network of human
acquaintances which is both large---containing over a million people---and
for which a precise definition of acquaintance is possible.  That network
is the network of scientific collaboration, as documented in the papers
which scientists write.

\section{Scientific collaboration networks}
We study networks of scientists in which two scientists are considered
connected if they have coauthored a paper together.  This seems a
reasonable definition of scientific acquaintance: most people who have
written a paper together will know one another quite well.  It is a
moderately stringent definition, since there are many scientists who know
one another to some degree but have never collaborated on the writing of a
paper.  Stringency however is not inherently a bad thing.  A stringent
condition of acquaintance is perfectly acceptable, provided, as in this
case, that it can be applied consistently.

We have constructed collaboration graphs for scientists in a variety of
fields.  The data come from four databases: MEDLINE (which covers published
papers on biomedical research), the Los Alamos e-Print Archive (preprints
primarily in theoretical physics), SPIRES (published papers and preprints
in high-energy physics), and NCSTRL (preprints in computer science).  In
each case, we examine papers which appeared in a five year window from 1995
to 1999 inclusive.  The sizes of the databases range from 2~million papers
for MEDLINE to $13\,000$ for NCSTRL.

The fact that some of the databases used contain unrefereed preprints
should not be regarded negatively.  Although unrefereed preprints may be of
lower average scientific quality than papers in peer-reviewed journals,
they are, as an indicator of social connection, every bit as good as their
refereed counterparts.

The idea of constructing a scientific collaboration network from the
publication record is not new, although no detailed study has previously
been published.  Among mathematicians the concept of the {\bf Erd\"os
  number} has long been current.  Paul Erd\"os was an influential, but
itinerant, Hungarian mathematician, who apparently spent a large portion of
his later life living out of a suitcase and writing papers with those of
his colleagues willing to give him room and board.  He published more
papers during his life than any other mathematician in history (at least
1400).  The Erd\"os number measures a mathematician's proximity, in
bibliographical terms, to the great man.  Those who have published a paper
with Erd\"os have an Erd\"os number of~1.  Those have published with a
coauthor of Erd\"os have an Erd\"os number of~2.  And so on.  An exhaustive
list exists of all mathematicians with Erd\"os numbers of~1
and~2~\cite{GI95}.

There are in addition many other interesting quantities to be measured on
collaboration networks, including the number of collaborators of
scientists, the numbers of papers they write, and the degree of
``clustering,'' which is the probability that two of a scientist's
collaborators have themselves collaborated.  All of these quantities and
several others are considered in this paper.

\begin{table*}
\setlength{\tabcolsep}{8.5pt}
\begin{center}
\begin{tabular}{l|c|cccc|c|c}
 &         & \multicolumn{4}{c|}{Los Alamos e-Print Archive}           &        \\
\cline{3-6}
 & MEDLINE & complete & {\tt astro-ph} & {\tt cond-mat} & {\tt hep-th} & SPIRES & NCSTRL \\
\hline
 total papers               & $2156769$    & $98502$     & $22029$     & $22016$     & $19085$     & $66652$       & $13169$       \\
 total authors              & $1388989$    & $52909$     & $16706$     & $16726$     & $8361$      & $56627$       & $11994$       \\
 \quad first initial only   & $1006412$    & $45685$     & $14303$     & $15451$     & $7676$      & $47445$       & $10998$       \\
 mean papers per author     & $5.5(4)$     & $5.1(2)$    & $4.8(2)$    & $3.65(7)$   & $4.8(1)$    & $11.6(5)$     & $2.55(5)$     \\
 mean authors per paper     & $2.966(2)$   & $2.530(7)$  & $3.35(2)$   & $2.66(1)$   & $1.99(1)$   & $8.96(18)$    & $2.22(1)$     \\
 collaborators per author   & $14.8(1.1)$  & $9.7(2)$    & $15.1(3)$   & $5.86(9)$   & $3.87(5)$   & $173(6)$      & $3.59(5)$     \\
 \quad cutoff $z_c$         & $7300(2700)$ & $52.9(4.7)$ & $49.0(4.3)$ & $15.7(2.4)$ & $9.4(1.3)$  & $1200(300)$   & $10.7(1.6)$   \\
 \quad exponent $\tau$      & $2.5(1)$     & $1.3(1)$    & $0.91(10)$  & $1.1(2)$    & $1.1(2)$    & $1.03(7)$     & $1.3(2)$      \\
 size of giant component    & $1193488$    & $44337$     & $14845$     & $13861$     & $5835$      & $49002$       & $6396$        \\
 \quad first initial only   & $892193$     & $39709$     & $12874$     & $13324$     & $5593$      & $43089$       & $6706$        \\
 \quad as a percentage      & $87.3(7)\%$  & $85.4(8)\%$ & $89.4(3)$   & $84.6(8)\%$ & $71.4(8)\%$ & $88.7(1.1)\%$ & $57.2(1.9)\%$ \\
 2nd largest component      & $56$         & $18$        & $19$        & $16$        & $24$        & $69$          & $42$          \\
 mean distance              & $4.4(2)$     & $5.9(2)$    & $4.66(7)$   & $6.4(1)$    & $6.91(6)$   & $4.0(1)$      & $9.7(4)$      \\
 maximum distance           & $21$         & $20$        & $14$        & $18$        & $19$        & $19$          & $31$          \\
 clustering coefficient $C$ & $0.072(8)$   & $0.43(1)$   & $0.414(6)$  & $0.348(6)$  & $0.327(2)$  & $0.726(8)$    & $0.496(6)$    \\
\end{tabular}
\end{center}
\caption{Summary of results of the analysis of seven scientific
  collaboration networks.  Numbers in parentheses are standard errors
  on the least significant figures.}
\label{results}
\end{table*}

\section{Results}
Table~\ref{results} gives a summary of some of the results of our analysis
of the databases described in the previous section.  In addition to results
for the four complete databases, we also give results for three
subject-specific subsets of the Los Alamos Archive, covering astrophysics
(denoted {\tt astro-ph}), condensed matter physics ({\tt cond-mat}), and
theoretical high-energy physics ({\tt hep-th}).  In this section we
highlight some of our results and discuss their implications.

\paragraph{Number of authors} Estimating the true number of distinct
authors in a database is complicated by two problems.  First, two authors
may have the same name.  Second, an author may identify themselves in
different ways on different papers, e.g.,~using first initial only, using
all initials, or using full name.  In order to estimate the size of the
error introduced by these effects, all analyses reported here have been
carried out twice.  The first time we use all initials of each author.
This will rarely confuse two different authors for the same person
(although this will still happen occasionally), but sometimes misidentifies
the same person as two different people, thereby overestimating the total
number of authors.  The second analysis is carried out using only the first
initial of each author, which will ensure that different publications by
the same author are almost always identified as such, but will with some
regularity confuse distinct authors for the same person.  Thus these two
analyses give upper and lower bounds on the number of authors, and hence
also of many other quantities that we are interested in.  In
Table~\ref{results} we quote both estimates of the number of authors for
each database.  For some other quantities we quote only an error estimate
based on the separation of the upper and lower bounds.

\paragraph{Mean papers per author and authors per paper} Authors typically
wrote about four papers in the five year period covered by this study.  The
average paper had about three authors.  Notable exceptions are in
theoretical high-energy physics and computer science in which smaller
collaborations are the norm (average two people), and the SPIRES
high-energy physics database with an average of 9 authors per paper.  The
reason for this last impressive figure is that the SPIRES database contains
data on experimental as well as theoretical work.  High-energy experimental
collaborations can run to hundreds or thousands of people, the largest
author list in the SPIRES database giving the names of a remarkable 1681
authors on a single paper.

\paragraph{Number of collaborators} The striking difference in
collaboration patterns in high-energy physics is further highlighted by the
results on the average number of collaborators of an author.  This is the
average total number of people with whom a scientist collaborates during
the period of study---the average degree, in the graph theorist's language.
For purely theoretical databases such as the {\tt hep-th} subset of the Los
Alamos Archive (covering high-energy physics theory) and NCSTRL (computer
science), this number is low, on the order of four.  For partly or wholly
experimental databases (condensed matter physics and astrophysics at Los
Alamos and MEDLINE (biomedicine)), the degree is significantly higher, as
high as 15 for astrophysics.  But high-energy experiment easily takes the
prize, with an average of 173 collaborators per author.

\begin{figure}
\begin{center}
\psfig{figure=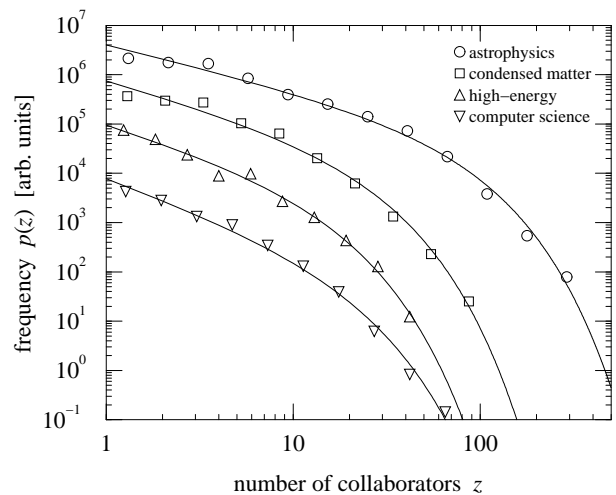,width=8cm}
\end{center}
\caption{Histograms of the number of collaborators of scientists in four
of the databases studied here.  The solid lines are least-squares fits to
Eq.~\eref{powerlaw}.}
\label{degree}
\end{figure}

There is more to the story of numbers of collaborators however.  In
Fig.~\ref{degree} we show histograms of the numbers of collaborators of
scientists in four of the smaller databases.  There has been a significant
amount of recent discussion of this distribution for a variety of networks
in the literature.  A number of authors~\cite{AJB99,BKMRRSTW00,FFF99} have
pointed out that if one makes a similar plot for the number of connections
(or ``links'') $z$ to or from sites on the World Wide Web, the resulting
distribution closely follows a power law: $P(z)\sim p^{-\tau}$, where
$\tau$ is a constant exponent with (in that case) a value of about $2.5$.
Barabasi and Albert have suggested~\cite{BA99} that a similar power-law
result may apply to all or at least most other networks of interest,
including social networks.  Others have presented a variety of evidence to
the contrary~\cite{ASBS00}.  Our data do not follow a power-law form
perfectly.  If they did, the curves in Fig.~\ref{degree} would be straight
lines on the logarithmic scales used.  However, our data are well fitted by
a power-law form with an exponential cutoff:
\begin{equation}
P(z) \sim p^{-\tau} \e^{-z/z_c},
\label{powerlaw}
\end{equation}
where $\tau$ and $z_c$ are constants.  Fits to this form are shown as the
solid lines in the figure.  In each case the fit has an $R^2$ of better
than $0.99$ and $p$-values for both power-law and exponential terms of less
than $10^{-3}$.

This form is commonly seen in physical systems, and implies an underlying
degree distribution which follows a power-law, but with some imposed
constraint that places a limit on the maximum value of~$z$.  One possible
explanation of this cutoff in the present case is that it arises as a
result of the finite (5~year) window of data used.  If this were the case,
we would expect the cutoff to increase with increasing window size.  But
even in the (impractical) limit of infinite window size, a cutoff would
still be imposed by the finite working lifetime of a professional scientist
(about 40 years).

The values of $\tau$ and $z_c$ are given in the table for each database.
The value of the cutoff size $z_c$, varies considerably.  For the mostly
theoretical condensed matter, high-energy theory, and computer science
databases it takes small values on the order of~10, indicating that
theorists rarely had more than this many collaborators during the five-year
period.  In other cases, such as SPIRES and MEDLINE it takes much larger
values.  In the case of SPIRES this is clearly again because of the
presence of very large experimental collaborations in the data.  MEDLINE is
more interesting.  A number of people have suggested to the author that the
presence of individuals with very large numbers of collaborators in the
biomedical community may be the result of the practice in that community of
laboratory directors signing their name to all papers emerging from their
laboratories.  One can well imagine that this would produce individuals
with a very high apparent number of collaborators.

The exponent $\tau$ of the power-law distribution is also interesting.  We
note that in all the ``hard sciences,'' this exponent takes values close
to~1.  In the MEDLINE (biomedicine) database however, its value is 2.5,
similar to that noted for the World Wide Web.  The value $\tau=2$ forms a
dividing line between two fundamentally different behaviors of the network.
For $\tau<2$, the average properties of the network are dominated by the
few individuals who have a large number of collaborators, while graphs with
$\tau>2$ are dominated by the little people---the ones with few
collaborators.  Thus, one finds that in biomedical research the highly
connected individuals do not determine the average characteristics of their
field, despite their names appearing on a lot of papers.  In physics and
computer science, on the other hand, it appears that they do.

\begin{figure}
\begin{center}
\psfig{figure=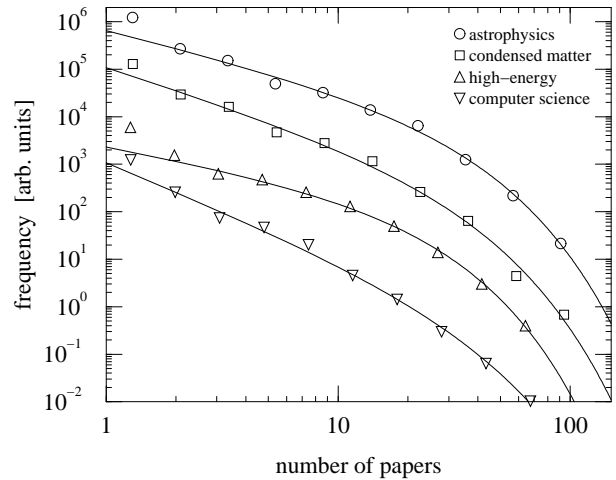,width=8cm}
\end{center}
\caption{Histograms of the number of papers written by scientists in four
  of the databases.  As with Fig.~\ref{degree}, the solid lines are
  least-squares fits to Eq.~\eref{powerlaw}.}
\label{papers}
\end{figure}

In Fig.~\ref{papers} we show histograms of the number of papers which
authors have written in the same four databases.  As the figure shows, the
distribution of papers follows a similar form to the distribution of
collaborators.  The solid lines are again fits to Eq.~\eref{powerlaw}, and
again match the data well in all cases.

\paragraph{The giant component} In all social networks there is the
possibility of a {\bf percolation transition}~\cite{SA91}.  In networks
with very small numbers of connections between individuals, all individuals
belong only to small islands of collaboration or communication.  As the
total number of connections increases, however, there comes a point at
which a {\bf giant component} forms---a large group of individuals who are
all connected to one another by paths of intermediate acquaintances.  It
appears that all the databases considered here are connected in this sense.
Measuring the size of groups of connected authors in each database, we find
(see table) that in most of the databases the largest such group occupies
around 80 or 90 percent of all authors: almost everyone in the community is
connected to almost everyone else by some path (probably many paths) of
intermediate coauthors.  In high-energy theory and computer science the
fraction is smaller but still more than half the total size of the network.
(These two databases may, it appears, give a less complete picture of their
respective fields than the others, owing to the existence of competing
databases with overlapping coverage.  The small size of the giant component
may in part be attributable to this.)

We have also calculated the size of the second-largest group of connected
authors for each database.  In each case this group is far smaller than the
largest.  This is a characteristic signature of networks which are well
inside the percolating regime.  In other words, it appears that scientific
collaboration networks are not on the borderline of connectedness---they
are very highly connected and in no immediate danger of fragmentation.
This is a good thing.  Science would probably not work at all if scientific
communities were not densely interconnected.

\paragraph{Average degrees of separation} We have calculated exhaustively
the minimum distance, in terms of numbers of intermediate acquaintances,
between all pairs of scientists in the databases studied.  We find that the
typical distance between a pair of scientists is about six; there are six
degrees of separation in science, just as there are in the larger world of
human acquaintance.  Even in very large communities, such as the biomedical
research community documented by MEDLINE, it takes an average of only six
steps to reach a randomly chosen scientist from any other, out of the more
than one million who have published.  We conjecture that this has a
profound effect on the way in which the scientific community operates.
Despite the importance of written communication in science as a document
and archive of work carried out, and of scientific conferences as a
broadcast medium for summary results, it is probably safe to say that the
majority of scientific communication takes place by private conversation.
Clearly, news of important discoveries and scientific information can
circulate far faster in a world where the typical separation of two
scientists is six, than it can in one where it is a thousand, or a million.

\begin{figure}
\begin{center}
\psfig{figure=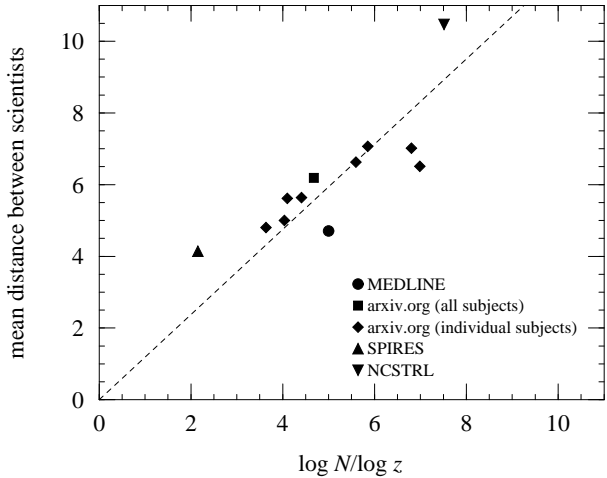,width=8cm}
\end{center}
\caption{Average distance between pairs of scientists in the various
communities, plotted against the average distance on a random graph of the
same size and average coordination number.  The dotted line is the best fit
to the data which also passes through the origin.}
\label{size}
\end{figure}

The variation of average vertex--vertex distances from one database to
another also shows interesting behavior.  The simplest model of a social
network is the {\bf random graph}---a network in which people are connected
to one another uniformly at random~\cite{Bollobas85}.  For a given number
$N$ of scientists with a given mean number $z$ of collaborators, the
average vertex--vertex distance on a random graph scales logarithmically
with $N$ according to $\log N/\log z$.  Social networks are measurably
different from random graphs~\cite{Watts99}, but the random graph
nonetheless provides a useful benchmark for comparing them against.  Watts
and Strogatz~\cite{WS98} defined a social network as being ``small'' if
typical distances were comparable with those on a random graph.  This
implies that such networks should also have typical distances which grow
roughly logarithmically in $N$, and indeed some authors
(e.g.,~Ref.~\onlinecite{ASBS00}) have used this logarithmic growth as the
defining criterion for a ``small world.''  In Fig.~\ref{size} we show the
average distance between all pairs of scientists for each of the networks
studied here, including separate calculations for the subject divisions of
the Los Alamos Archive, of which there are nine.  In total there are 12
points, which we have plotted against $\log N/\log z$, using the
appropriate values of $N$ and $z$ from Table~\ref{results}.  As the figure
shows, there is a strong correlation ($R^2=0.83$) between the measured
distances and the expected $\log N$ behavior, indicating that distances do
indeed scale logarithmically with the number of scientists in a community.

We also quote in Table~\ref{results} figures for the maximum separation of
pairs of scientists in each database, which tells us the greatest distance
we will ever have to go to connect two people together.  This quantity is
often referred to as the {\bf diameter} of the network.  For all the
networks examined here, it is on the order of~20; there is a chain of at
most about 20 acquaintances connecting any two scientists.  (This result of
course excludes pairs of scientists who are not connected at all, as will
often be the case for the 10 or 20 percent who fall outside the giant
component.)

\paragraph{Clustering} Watts and Strogatz~\cite{WS98} have pointed out
another important property of social networks which is absent from many
network models which have been employed by social scientists and
epidemiologists.  Real networks are {\bf clustered}, meaning that they
possess local communities in which a higher than average number of people
know one another.  A laboratory or university department might form such a
community in science, as might the set of researchers who work in a
particular sub-field.  Watts and Strogatz also proposed a way of probing
for the existence of such clustering in real network data.  They defined a
{\bf clustering coefficient} $C$, which for our purposes is the average
fraction of pairs of a person's collaborators who have also collaborated
with one another.  Thus for example a person with $z=10$ collaborators has
$\frac12z(z-1)=45$ pairs of collaborators.  If, say, 20 of those pairs have
also collaborated on a paper, then the clustering coefficient for that
person is $20/45=0.44$.  The same quantity calculated for the entire
network is the quantity we call~$C$.  Values are given in
Table~\ref{results}, and show that there is a very strong clustering effect
in the scientific community: two scientists typically have a 30 percent or
greater probability of collaborating if they have both collaborated with
another third scientist.  A number of explanations of this result are
possible.  To some extent it is certainly the result of the appearance of
papers with three or more authors: such papers clearly contain trios of
scientists who have all collaborated with one another.  However, the values
measured here cannot be entirely accounted for in this way, and indicate
also that scientists either introduce their collaborators to one another,
thereby engendering new collaborations, or perhaps that institutions bring
sets of collaborators together to form a variety of new collaborations.

The MEDLINE database is interesting in that it possess a much lower value
of the clustering coefficient than the ``hard science'' databases.  This
appears to indicate that it is significantly less common in biological
research for scientists to broker new collaborations between their
acquaintances than it is in physics or computer science.

\section{Conclusions}
We have analyzed the collaboration networks of scientists from biology and
medicine, various sub-disciplines of physics, and computer science, using
the author attributions from papers or preprints appearing in those areas
over a five year period from 1995 to 1999.  We find a number of interesting
properties of these networks.  In all cases, scientific communities seem to
constitute a ``small world'' in which the average distance between
scientists via a line of intermediate collaborators scales logarithmically
with the size of the relevant community.  Typically we find that only about
five or six steps are necessary to get from one randomly chosen scientist
in a community to another.  We conjecture that this smallness is a crucial
feature of a functional scientific community.

We also find that the networks are highly clustered, meaning that two
scientists are much more likely to have collaborated if they have a third
common collaborator than are two scientists chosen at random from the
community.  This may indicate that the processes of scientists introducing
their collaborators to one another is an important one in the development
of scientific communities.

We have studied the distributions of both the number of collaborators of
scientists and the numbers of papers they write.  In both cases we find
these distributions are well fit by power-law forms with an exponential
cutoff.  This cutoff may be due to the finite time window used in the
study.

We find a number of significant statistical differences between different
scientific communities.  Some of these are obvious: experimental
high-energy physics, for example, which is famous for the staggering size
of its collaborations, has a vastly higher average number of collaborators
per author than any other field examined.  Other differences are less
obvious however.  Biomedical research, for example, shows a much lower
degree of clustering than any of the other fields examined.  In other words
it is less common in biomedicine for two scientists to start a
collaboration if they have another collaborator in common.  Biomedicine is
also the only field in which the exponent of the distribution of numbers of
collaborators is greater than~2, implying that the average properties of
the collaboration network are dominated by the many people with few
collaborators, rather than, as in other fields, by the few people with
many.

The work reported in this paper represents, inevitably, only a first look
at the collaboration networks described.  Many theoretical measures have
been described, in addition to the distances and clustering studied here,
which reflect socially important structure in such networks.  We hope that
academic collaboration networks will prove a reliable and copious source of
data for testing out such theories, as well as being interesting in their
own right, especially to ourselves, the scientists whom they describe.

\section*{Acknowledgements}
The author is indebted to Paul Ginsparg and Geoffrey West (Los Alamos
e-Print Archive), Carl Lagoze (NCSTRL), Oleg Khovayko, David Lipman
and Grigoriy Starchenko (MEDLINE), and Heath O'Connell (SPIRES), for
making available the publication data used for this study.
The author would also like to thank Dave Alderson, Paul Ginsparg, Laura
Land\-weber, Steve Strogatz, and Duncan Watts for illuminating
conversations.  This work was funded in part by a grant from Intel
Corporation to the Santa Fe Institute Network Dynamics Program.  The NCSTRL
digital library was made available through the DARPA/CNRI test suites
program funded under DARPA grant N66001--98--1--8908.

\def\refer#1#2#3#4#5#6{{\frenchspacing\sc#1,}\hspace{8pt}#2
                       {\frenchspacing#3} {\bf#4}, #5 (#6).}
\def\bookref#1#2#3#4{{\frenchspacing\sc#1,} {\it#2} #3 (#4).}

\end{document}